\PassOptionsToPackage{hyphens}{url}
\PassOptionsToPackage{svgnames,dvipsnames,x11names}{xcolor}
\documentclass[11pt]{article}

\usepackage[letterpaper,margin=1in]{geometry}
\usepackage[T1]{fontenc}
\usepackage[utf8]{inputenc}
\usepackage{microtype}
\usepackage{amsmath,amssymb,amsthm,bm}
\usepackage{cite}
\usepackage{xcolor}
\usepackage{enumitem}
\usepackage{graphicx}
\usepackage{tabularx}
\usepackage{booktabs}
\usepackage{multirow}
\usepackage{placeins}
\usepackage{wrapfig}
\usepackage{tikz}
\usetikzlibrary{positioning,arrows.meta,calc,backgrounds,shapes.geometric,fit}
\usepackage{pgfplots}
\pgfplotsset{compat=1.18}
\usepackage{listings}
\usepackage{xspace}
\usepackage{tcolorbox}
\usepackage[normalem]{ulem}
\usepackage{caption}
\usepackage{subcaption}
\usepackage{titlesec}
\usepackage{fancyhdr}
\usepackage{hyperref}
\hypersetup{
    colorlinks=true,
    linkcolor=blue!50!black,
    citecolor=blue!50!black,
    urlcolor=blue!50!black,
    pdftitle={RepoRescue: Empirical Study of LLM Agents on Whole-Repository Compatibility Rescue},
}

\titleformat{\section}{\Large\bfseries}{\thesection}{0.7em}{}
\titleformat{\subsection}{\large\bfseries}{\thesubsection}{0.7em}{}
\titleformat{\subsubsection}{\normalsize\bfseries}{\thesubsubsection}{0.7em}{}
\titlespacing*{\section}{0pt}{2.1ex plus 0.4ex minus 0.2ex}{1.0ex}
\titlespacing*{\subsection}{0pt}{1.6ex plus 0.3ex minus 0.2ex}{0.7ex}
\titlespacing*{\subsubsection}{0pt}{1.2ex plus 0.2ex minus 0.1ex}{0.5ex}

\definecolor{passgreen}{RGB}{0,140,0}
\definecolor{failred}{RGB}{200,30,30}
\definecolor{warnorg}{RGB}{200,120,0}
\definecolor{lstkeyword}{RGB}{0,80,180}
\definecolor{lstcomment}{RGB}{110,110,110}
\definecolor{lststring}{RGB}{34,100,34}

\newcommand{\reporescue}{\textsc{RepoRescue}}

\newcommand{\reproref}[1]{}

\newcommand{\findingbox}[2]{\begin{tcolorbox}
[boxsep=2pt, left=4pt, right=4pt, top=1pt, bottom=1pt,
 colback=yellow!8, colframe=brown!40, coltext=black, title={#1}]
{#2}
\end{tcolorbox}}

\lstdefinestyle{pystyle}{
  language=Python,
  basicstyle=\ttfamily\footnotesize,
  keywordstyle=\color{lstkeyword}\bfseries,
  commentstyle=\color{lstcomment}\itshape,
  stringstyle=\color{lststring},
  numbers=left,
  numberstyle=\tiny\color{gray},
  numbersep=5pt,
  frame=single,
  framesep=3pt,
  showstringspaces=false,
  breaklines=true,
  tabsize=2,
  morekeywords={with,as,async,await}
}

\title{RepoRescue: An Empirical Study of LLM Agents on Whole-Repository Compatibility Rescue}

\author{}
\date{}

\makeatletter
\renewcommand{\maketitle}{%
  \begin{center}
    \vspace*{-1.0em}
    {\LARGE\bfseries \@title\par}
    \vspace{0.9em}
    {\large
    Zhihao Lin\textsuperscript{1}\quad
    Mingyi Zhou\textsuperscript{1}\quad
    Zhensu Sun\textsuperscript{2}\quad
    Yizhuo Yang\textsuperscript{1}\par
    \vspace{0.2em}
    Renyu Yang\textsuperscript{1}\quad
    David Lo\textsuperscript{2}\quad
    Li Li\textsuperscript{1,*}\par}
    \vspace{0.65em}
    {\small
    \textsuperscript{1}Beihang University, Beijing, China\qquad
    \textsuperscript{2}Singapore Management University, Singapore\par}
    \vspace{0.35em}
    {\small
    \{mathieulin, zhoumingyi, yangyizhuo, renyuyang\}@buaa.edu.cn,
    lilicoding@ieee.org\par
    zhensuuu@gmail.com,\ davidlo@smu.edu.sg\par}
    \vspace{0.25em}
    {\small \textsuperscript{*}Corresponding author.\par}
  \end{center}
  \vspace{0.8em}
  \noindent\rule{\textwidth}{0.4pt}
  \vspace{0.9em}
}
\makeatother

\renewenvironment{abstract}
  {\begin{tcolorbox}[
    colback=gray!5,
    colframe=gray!35,
    boxrule=0.4pt,
    arc=2pt,
    left=10pt,
    right=10pt,
    top=8pt,
    bottom=8pt]
   \small\noindent\textbf{Abstract.}\enspace}
  {\end{tcolorbox}\vspace{0.6em}}

\begin{document}

\maketitle

\begin{abstract}
Open-source libraries and tools are widely reused, but compatibility maintenance is expensive. Once maintainers leave, otherwise useful repositories can stop working as runtimes and dependencies evolve. We study whether LLM agents can perform this form of maintenance: adapting old repositories so that they work on modern environments.
We call this task \emph{compatibility rescue}. Unlike bug repair, where a program violates its intended behavior in the environment for which it was written, compatibility rescue starts from a repository that still works in its original environment and then fails after the runtime or dependency ecosystem changes. \reporescue{} gives the agent only the repository and its failing modern environment. The agent must diagnose the failure, locate the affected code, and produce a source-code rescue that restores the whole historical suite.
We build \reporescue{} from 193 Python and 122 Java repositories, each checked to pass in its historical environment and fail after modernization before agents attempt a rescue. We run five deployed agent systems on Python and three on Java. Pass rate tells us whether the test suite was restored. To check whether the patch repaired source code, we also rerun each submitted patch after removing test-file edits; we call this \emph{source-only evaluation} because it asks whether the remaining source changes alone restore the suite. We further add a runtime-enforced regime that blocks test edits during the session and practical-use validation for repositories whose original suites pass after rescue.
We report four findings:
(1) All four Claude Code systems sometimes edit failing tests even when the prompt forbids it; when test edits are blocked at runtime, Kimi still rescues 41.5\% of repositories. (2) The systems succeed on different repositories, and their union (62.7\%) exceeds the best single system (51.8\%) by 10.9 percentage points.
(3) Difficulty tracks the amount of cross-file reasoning required. On 14 repositories that need coordinated whole-codebase changes, GPT-5.2 through Codex is recorded as passing all 14, while every Claude Code system passes at most 2; the traces point to gaps in planning and coordination.
(4) A passing test suite is only a first signal: among 34 unmaintained Python candidates whose original suites pass after rescue, 22 work in realistic scenarios, and 12 pass bug-hunt with patches that address the compatibility failure.
As a benchmark, \reporescue{} measures these capabilities jointly and labels each rescue on a reasoning-level hierarchy, from mechanical edits to whole-codebase coordination.
\end{abstract}

% Sec 1: Introduction
% ==========================================
% 1. Introduction
% ==========================================
\section{Introduction}
\label{sec:intro}

Software often outlives its maintainers. Libraries can stop receiving releases for many reasons: maintainers move on, funding ends, or the project becomes stable enough that no one keeps updating it. The code may still be useful, but the environment around it does not stand still because Python, Java, build tools, and dependencies continue to change. Over time, a project that worked in its original environment can lose compatibility with the current one, and new users can no longer import, build, or reuse it.

\begin{figure*}[t]
\centering
\includegraphics[width=\textwidth]{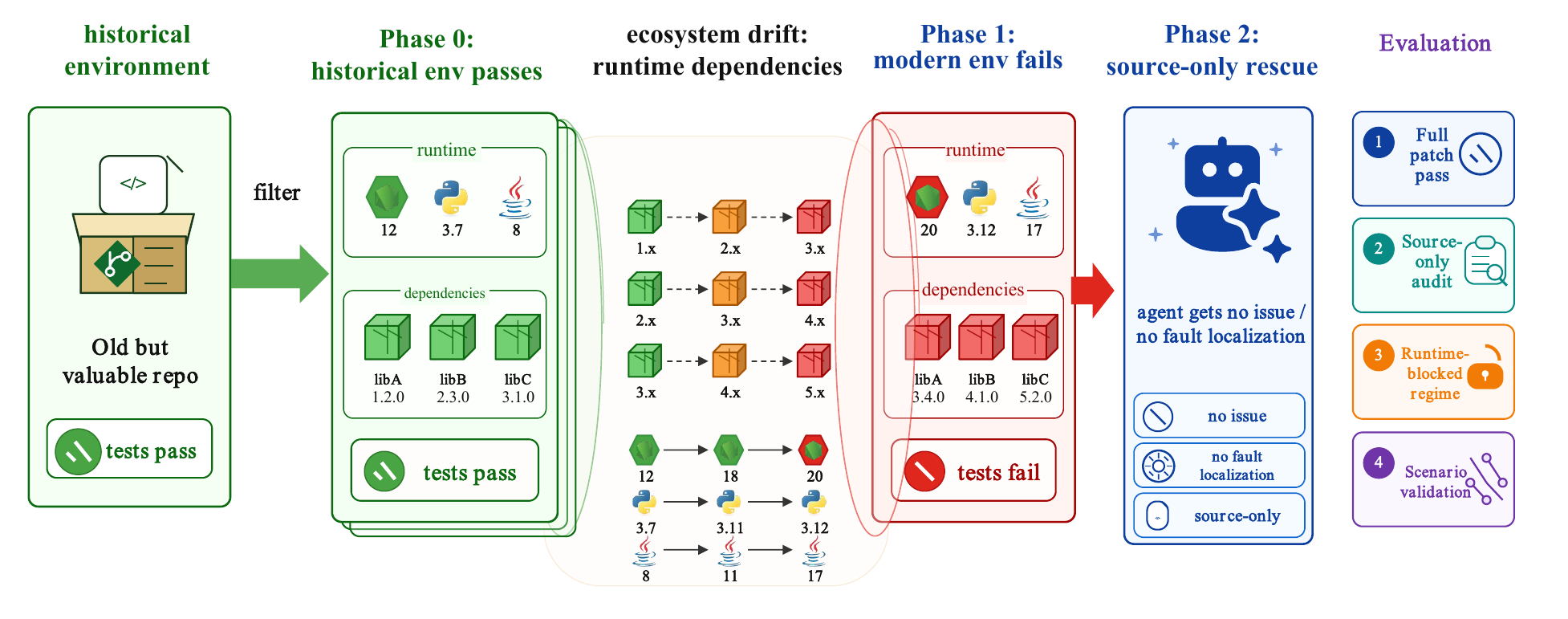}
\caption{Overview of \reporescue{}. We admit repositories that pass in a historical environment (Phase~0) and fail after ecosystem drift (Phase~1), then ask an agent to produce a source-only rescue (Phase~2). We evaluate each outcome through full-patch pass, source-only audit, runtime blocking, and realistic scenario validation.}
\label{fig:overview}
\end{figure*}

These projects can remain useful after maintenance stops. Across 47 unmaintained but still-depended-on Python libraries, we found 2{,}851 forks created after maintenance stopped. That number shows downstream demand, and it also shows fragmentation: no single fork replaced the original maintainer, and downstream consumers were left choosing among partially maintained copies~\cite{coelho2017why,avelino2019abandonment}. If such projects could be kept compatible with modern environments, prior engineering effort could remain available to more developers instead of being repeatedly reimplemented or patched downstream.
In this paper, we call this task \emph{compatibility rescue}: adapting historically working software to a modern runtime or dependency environment without changing its intended behavior.
Figure~\ref{fig:overview} summarizes the setting: we first establish that an old repository worked, then confirm that ecosystem drift breaks it, and finally ask an agent to restore source compatibility.
Compatibility rescue is different from general bug repair. Its object is historical software that once worked, then lost compatibility as runtimes and dependencies changed. The maintenance problem is to adapt that working software to a modern environment without changing its intended behavior; \S\ref{sec:motivation} formalizes the task and its boundary with adjacent forms of repair.

At ecosystem scale, this maintenance work is expensive. A maintainer has to recover an old environment, reproduce the breakage under a modern one, and then change the source while preserving the library's behavior.
Developers can use tools such as \texttt{pyupgrade}~\cite{pyupgrade} and OpenRewrite~\cite{openrewrite} to automate part of this work. Their coverage is strongest for predefined migration patterns, while many broken repositories require project-specific diagnosis.
LLM agents create a different possibility because compatibility adaptation often requires more than syntactic rewrites: an agent may need to run tests, inspect dependency source, read changelog-shaped failures, propagate API substitutions across files, and decide whether source changes restore the intended behavior. Existing agent studies have not yet measured whole-repository adaptation from only a failing modern environment, or how agent behavior changes when shortcuts such as editing tests are unavailable.

To fill this gap, we build \reporescue{}, a benchmark and empirical study of LLM agents on whole-repository compatibility rescue. The benchmark contains 193 Python repositories and 122 Java repositories. Each task is admitted only after the repository passes in its historical environment and fails after modernization; agents then try to restore the original suite through source-code changes. The detailed dataset construction, including unmaintained projects and time-travel snapshots, is in \S\ref{sec:dataset}.

The benchmark size reflects this admission standard rather than raw repository availability. For Python, a candidate must show reuse signal, survive manual removal of forks, mirrors, and demos, pass its original unmodified suite in a reconstructed historical environment, and fail deterministically only after modernization. This filter shrinks the unmaintained track from 213 already-filtered candidates to 47 validated rescue subjects. We then add 146 time-travel snapshots only when a maintainer's subsequent compatibility fix gives additional ground truth. For Java, a separate Maven filter starts from 232 dormant candidates; historical and modern environment checks plus build-configuration normalization leave 122 repositories that still require source-code changes. \reporescue{} therefore trades raw scale for high-confidence tasks in which the starting state, failure trigger, and rescue target are all testable.

Using this benchmark, we ask how far deployed LLM agents can go, where their successes complement or diverge, what makes a rescue hard, and whether a restored test suite is enough for practical reuse.
\textbf{RQ1} asks whether deployed agents can rescue compatibility failures. Across 193 Python repositories, full-patch pass rates reach 36.8--51.8\%, but source-only auditing lowers the four Claude Code systems to 19.7--24.4\%, while GPT-5.2 through Codex retains 49.7\%. Blocking test edits during the run changes behavior: Kimi still rescues 41.5\% of repositories.
\textbf{RQ2} asks whether systems solve the same repositories. The five-system union is 10.9~pp above the best single system, so the benchmark points to routing and portfolio questions.
\textbf{RQ3} asks what makes rescues hard. Difficulty is concentrated in coordinated whole-codebase repairs: GPT-5.2 through Codex is recorded as passing all 14 L4 repositories, while every Claude Code system passes at most 2.
\textbf{RQ4} asks whether passing the original suite is enough for reuse. Among 34 unmaintained Python candidates whose original suites pass after rescue, 22 work in realistic scenarios and 12 pass bug-hunt with patches that address the compatibility failure. The Java track adds a related caution: in 6 repositories, test edits damage otherwise working source.

This paper makes three contributions:
\begin{enumerate}[leftmargin=*, itemsep=2pt, topsep=2pt]
    \item \textbf{A benchmark for compatibility rescue.} \reporescue{} contains 193 Python repositories (47 unmaintained and 146 time-travel, the latter carrying maintainer ground-truth fixes) and 122 unmaintained Java repositories. Its validation protocol checks that each task starts from historically working code, fails after the runtime or dependency environment is modernized, and is then evaluated under the original test command.
    \item \textbf{An empirical study of deployed LLM agents.} We run 965 Python primary trials, 386 Python enforced re-runs, and 366 Java trials to measure how often agents restore compatibility, how much apparent success depends on test edits, and where different systems solve complementary repositories.
    \item \textbf{An analysis of rescue difficulty and practical usability.} We label successful repairs by reasoning level (L1--L4, $\kappa$=0.76), validate 34 unmaintained Python rescues beyond the original suite, and analyze 108 Java rescue outcomes to show how static typing exposes shortcut harm.
\end{enumerate}

% Sec 2: Compatibility rescue as a benchmark task
% ==========================================
% 2. Compatibility Rescue as a Benchmark Task
% ==========================================
\section{Compatibility Rescue}
\label{sec:motivation}

In our setting, a repository qualifies for compatibility rescue only if it once passed its own tests in an original environment but fails after the runtime or dependency ecosystem is modernized. A valid rescue restores compatibility in the modern environment while preserving the behavior encoded by the historical test suite. This separates rescue from bug repair, which fixes a defect in the environment for which the program was written, and from project build repair, which may legitimately change dependency specifications or build scripts to make a repository runnable. \reporescue{} focuses instead on version adaptation of historically working code.

The need is practical because old but still-used libraries break across source, tests, and dependency specifications as their base runtime moves forward. 
For example, Python~3.13 removes \texttt{cgi} and \texttt{distutils}, while NumPy~2 drops legacy type aliases. Java has analogous pressure from JDK~21's module system and the \texttt{javax}-to-\texttt{jakarta} move. These changes leave the libraries' intended behavior intact while removing assumptions once supplied by the environment. Many such libraries still sit on active dependency paths: \texttt{requests-html}, for instance, has had no release since 2019 yet still draws over a million PyPI downloads per month.
This also matters for agentic software engineering. Many old developer tools remain useful, and today they can be exposed to agents through a thin MCP~\cite{mcp} wrapper. The wrapper still depends on the underlying library running on a modern runtime. Later, we use PyCG~\cite{pycg2021}, wrapped behind FastMCP~\cite{fastmcp}, as one example: the historical suite can pass while the main downstream call path still fails.

Existing deterministic modernizers cover the regular syntactic part of this space. Tools such as \texttt{pyupgrade}~\cite{pyupgrade} and OpenRewrite's \texttt{UpgradeToJava21}~\cite{openrewrite} rewrite well-defined patterns; compatibility rescue also requires dependency-source inspection, changed runtime contracts, and coordinated semantic edits across files.
Compatibility rescue gives LLM agents a benchmark setting centered on historical software, suite-wide failure, and runtime or dependency drift. The benchmark must establish three facts: the repository once worked, modernization breaks the same repository, and a passing repair reflects source adaptation. The rest of the paper makes these facts observable through historical and modern environment validation, source-only auditing, and realistic-use checks for rescues whose historical suite passes.

% Sec 3: Methodology (includes dataset, validation, regimes, RQs)
% ==========================================
% 3. Methodology
% ==========================================
\section{Benchmark Construction and Evaluation}
\label{sec:benchmark}

\subsection{Dataset Construction}
\label{sec:dataset}

To instantiate the task in \S\ref{sec:motivation}, we build a benchmark, \reporescue{}, which comprises 193 Python repositories and 122 Java repositories.
% \su{The structure of this subsection could be an overview, a step-by-step explanation about how this dataset is built, and the evaluation metrics. In the overview, we introduce the statistics of the dataset. In the steps, we start from how the raw data is filtered and then validated. Phase~0~PASS and Phase~1~FAIL should not appear at beginning since they are introduced at the end.
We filter repositories through two environment checks. We first construct an original environment for each candidate with agent assistance, because the rescue task only makes sense when the project can still be shown to work in its own historical setting. We keep the candidate only if its unmodified test suite passes in that environment. We then move the same repository to a modern environment, using Python~3.13 or JDK~21 with current dependencies. We keep it only if the same suite now fails for a compatibility reason. We call these two checks Phase~0 and Phase~1, and use them to define the benchmark subjects.

\textbf{Python: unmaintained repositories (47).}
We collect these repositories from GitHub using filters meant to capture projects with both downstream value and long-term dormancy. Concretely, a candidate must have at least 100~stars, no commits and no Python~3.10--3.13 pull requests for at least 24~months, a last release before Python~3.10 (October~2021), and a non-archived status. These filters yield 213 candidates. Manual inspection removes forks, mirrors, and demonstration projects, leaving 94 repositories. We then construct and freeze the Phase~0 and Phase~1 environments for each candidate with agent assistance. After rerunning each repository's original test command in the frozen environments, 47 repositories remain, each passing in the historical environment and failing after modernization.

\textbf{Python: time-travel (146).}
The unmaintained track gives only 47 high-quality candidates after the GitHub filters, manual inspection, and Phase~0 and Phase~1 environment checks. Python's fast-moving dependency ecosystem often breaks a dormant suite even under its original interpreter, so the suite no longer passes Phase~0. We therefore broaden the benchmark with active repositories that expose the same kind of compatibility breakage. We harvest 260 of them, scan each project's history for a commit in which the maintainer fixed a compatibility problem, and check out the commit \emph{immediately before} that fix. This pre-fix snapshot breaks on a modern runtime in the same way an unmaintained project would. After the same Phase~0 and Phase~1 checks, 146 snapshots remain. Unlike the unmaintained set, each snapshot comes with the maintainer's own subsequent fix, which we keep as ground truth.

\textbf{Java: 122 unmaintained repositories.}
For Java, we use a separate GitHub filter for Maven-based projects with at least 10~stars and no commits for at least 12~months. This yields 232 candidates. After the Phase~0 and Phase~1 checks, 192 candidates remain. Many Java failures, however, stem from aged build configuration rather than source code, so we first normalize this layer uniformly. We bump \texttt{source} and \texttt{target} levels, upgrade plugins, add \texttt{--add-opens}, and migrate \texttt{javax} to Jakarta. This normalization is applied before task admission and is not counted as an agent rescue action; its purpose is to remove aged build configuration as the dominant failure mode. After normalization, 122 repositories still break and thus require source-code modification, which are the ones we keep. This upfront step isolates source-level faults and makes the compile-versus-runtime distinction clean. Among the 122 repositories, Phase~1 failures split into 52 compilation errors and 70 runtime or test failures.

\textbf{Dataset summary and deterministic baselines.}
The 193 Python repositories exercise roughly 68{,}895 Phase~0 tests, with a median of 165 tests per repository. They span Python~2.7 through~3.13, with Python~3.10 the most common, and contain 2 to 1{,}847 source files each. Their main breakage causes are dependency API changes (113~repositories), standard-library module removals (40), and standard-library API removals (27). The Java track is dominated by API removal and tightened reflection. As simple deterministic baselines, \texttt{pyupgrade}~\cite{pyupgrade} rescues 28 of 193 Python repositories (14.5\%), and OpenRewrite's \texttt{UpgradeToJava21}~\cite{openrewrite} rescues 3 of 122 Java repositories (2.5\%) under the same source-only scoring.

\subsection{Validation Protocol}
\label{sec:validation}

\textbf{Recovering the historical environment (Phase~0).} An author rebuilds the environment each repository originally ran in, one repository at a time, with Codex only as an assistant. From the project's own lockfile, or from a PyPI snapshot bounded by its last-commit date plus its documented build steps, we assemble the environment with \texttt{uv} and pin the era-matched interpreter. We admit a repository only when its own unmodified test suite passes; that test result is the evidence of a recovered working state. We then freeze the \texttt{site-packages} for reproducibility. Since admission is decided by the suite, not by any model, and the frozen environment is later handed identically to every system at Phase~2, this assistance advantages no system at adaptation time.

\textbf{Exposing modern breakage (Phase~1).} We move the same repository to a modern environment. For Python, this is a fresh Python~3.13 virtual environment with no version pins, so the suite runs against current dependencies. For Java, this is the JDK~21 environment described above. Every Phase~1 failure is re-validated on a clean rebuild to confirm that it is deterministic.

\textbf{Evaluating the rescue (Phase~2).} We hand the agent the pre-built Phase~1 environment and the repository tree. The task input contains the failing project state, with no issue description, fault localization, or incompatibility label. The agent diagnoses by running the tests, reads the installed dependencies' source on disk to discover changed APIs, and edits the repository's own source. The harness forbids test-file edits, dependency-specification edits, and package installation through \texttt{pip} or \texttt{mvn install}. Phase~2 re-runs the project's historical test command verbatim, including any pre-existing \texttt{--ignore} flags or test-path scoping; this keeps the test surface aligned with what the maintainer originally configured. A repository passes Phase~2 only if the suite reports no failures and the passing-test count is within 5\% of Phase~0, on both Python and Java.

\subsection{Evaluation Protocol}
\label{sec:evaluation}

We use source-only evaluation operationally. After an agent finishes, we remove any edits to test files from its submitted patch and rerun Phase~2; this asks whether the remaining source-code changes are sufficient to restore the historical suite. The Phase~2 task instruction also forbids dependency-specification edits and package installation, but agents may still attempt forbidden actions in the soft-constraint setting. \emph{Full-patch} evaluation reruns the suite with every submitted edit and therefore measures submitted-patch suite restoration. \emph{Post-hoc source-only} evaluation is the test-edit-removal rerun described above. \emph{Runtime-blocked source-only} evaluation prevents test writes, dependency-specification edits, and package installation during the session, asking how the same agent behaves when the shortcut path is unavailable. We use the runtime-blocked setting as a targeted ablation on Kimi~K2.5 and GLM-5, two systems that share the Claude Code framework but show different post-hoc shortcut patterns in \S\ref{sec:gap}.

\subsection{Validation Beyond the Historical Suite}
\label{sec:scenario}
Passing Phase~2 means that the original suite is green again. We treat that as suite restoration, then ask whether the repaired package is still usable outside that suite. For each Phase~2-passing unmaintained Python candidate, we first audit whether the repository actually needed rescue and whether the patch changed source code related to the Python~3.13 failure. We then install the rescued package in a clean virtual environment and exercise realistic entry points. For reproducibility, each scenario script imports and exercises at least three public submodules when available, includes at least one call path related to the Phase~1 break surface, and asserts returned values, raised exceptions, or observable side effects rather than merely checking imports. When a maintained downstream package or wrapper exists, we also run a small downstream scenario against the rescued library. Targeted bug-hunt probes then check the behavior touched by the rescue patch, so we can count rescue-caused regressions separately from upstream or pre-existing issues.

\section{Methodology}
\label{sec:method}

\subsection{Research Questions}
\label{sec:rqs}

We organize the study around four questions that determine whether agent-based compatibility rescue is useful in practice. \textbf{RQ1.} Can deployed agents rescue compatibility failures in real repositories? \textbf{RQ2.} Do agent systems solve the same repositories, or do their successes complement each other? \textbf{RQ3.} What makes a rescue task hard for agents? \textbf{RQ4.} When the original tests pass, does the rescued library work in realistic use? Together, these questions move from rescue capability, to cross-system complementarity, to task difficulty, and finally to usability beyond the original test suite.

\subsection{Agent Systems}
\label{sec:systems}
We use \emph{system} to mean an LLM paired with an agent framework, and treat that pair as the unit of behavioral observation. The distinction matters because the harness can affect context construction, tool use, retries, and stopping decisions; recent work argues that agent comparisons can misattribute harness effects to backend models when the harness is not disclosed~\cite{zhang2026harnessdisclosure}. We therefore report behavior at the system level before discussing model-side or framework-side mechanisms.

Our primary comparison stays within one framework. The four Claude Code systems share the same harness (prompts, tool schema, retry logic), so differences among them approximate model differences, though provider-side sampling defaults remain a residual confound. GPT-5.2 runs through the Codex framework. We treat its result as a cross-framework observation throughout and report the framework as one candidate mechanism for its gap rather than crediting the model alone.

On the Python track we run Claude Code CLI~\cite{anthropic2024claude} with Claude~Sonnet~4.6, GLM-5, Kimi~K2.5, or MiniMax~M2.5, together with GPT-5.2 through Codex CLI~\cite{chen2021codex}. On the Java track we run GPT-5.2 through Codex, GLM-5 through Claude Code, and Kimi~K2.5 through Claude Code.

\subsection{Implementation Details}
\label{sec:implementation}

Each repository$\times$system pair is a single trial, giving 965 primary Python trials, 366 Java trials, and 386 runtime-enforced re-runs, for 1{,}717 in total. We report rates as observations under this deployment snapshot, each with a 95\% Wilson confidence interval~\cite{wilson1927ci}, and group repositories into difficulty tiers by how many systems pass them: Easy (at least four), Medium (one to three), and Hard (none).

% Sec 4: Results (RQ1-4 + Java)
% ==========================================
% 4. Results
% ==========================================
\section{Results}
\label{sec:results}

%The results are not a single leaderboard.
On Python, the four checks separate different behaviors: source-only and enforced scoring distinguish source repairs from test-edit shortcuts~(RQ1); union and intersection counts measure complementarity~(RQ2); reasoning levels locate coordination failures~(RQ3); and post-PASS validation tests what Phase~2 leaves untested~(RQ4). We then use Java as a cross-ecosystem extension of RQ1 and RQ2, because compilation failures and runtime failures can be separated more cleanly there. All rates are single-trial estimates for the deployed-system snapshot in \S\ref{sec:method}. We evaluate model--framework pairings in their practical form: GPT-5.2 through Codex, Sonnet~4.6 through Claude Code, and GLM-5, Kimi~K2.5, and MiniMax~M2.5 in the shared Claude Code framework. Since the framework affects context construction, tools, retries, and stopping behavior, our comparisons stay at the system-behavior level.

% ------------------------------------------
% RQ1: Constraint Compliance under Post-hoc and Enforced Audits
% ------------------------------------------

\subsection{RQ1: Can deployed agents rescue compatibility failures in real repositories?}
\label{sec:gap}

Agents can rescue compatibility failures, but a raw pass rate is too broad: a patch may pass because it fixes the source, or because it changes the tests. We therefore use source-only success (success after removing test-file edits) as the main capability measure, following SWE-bench's convention of evaluating a patch after removing inadmissible test edits~\cite{jimenez2024swebench}. The gap between full-patch and source-only measures dependence on test edits. The post-hoc audit strips those edits after the run; the runtime-enforced ablation, run for Kimi~K2.5 and GLM-5, blocks the shortcut during the session and asks whether the agent chooses a different repair.

% Figure 2: The gap between full-patch, source-only, and enforced pass rates
\begin{figure}[t]
\centering
\includegraphics[width=\columnwidth]{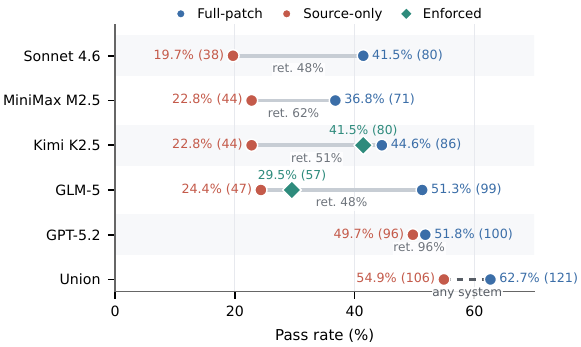}
\caption{Python rescue outcomes on 193 repositories. Sonnet, MiniMax, Kimi, and GLM-5 run through Claude Code; GPT-5.2 runs through Codex. Each row connects full-patch success (blue) to post-hoc source-only success (red); marker labels report pass rate with passing-repository count in parentheses. The inline ``ret.'' label is source-only retention relative to full-patch success. Green diamonds show runtime-blocked results for Kimi and GLM-5. Wilson confidence intervals~\cite{wilson1927ci} are reported in the artifact.}
\label{fig:gap}
\end{figure}

Figure~\ref{fig:gap} quantifies the gap. The four Claude Code systems reach 36.8--51.3\% full-patch success, but source-only scoring lowers them to 19.7--24.4\%. The 14--27\% drop means that 38--53\% of their apparent successes depend on forbidden test edits. GPT-5.2 through Codex behaves differently, retaining 96\% of its full-patch successes under the same audit. Within the Claude Code group, much of the full-patch spread therefore comes from shortcut frequency rather than a clean capability separation, although provider-side sampling defaults remain a residual confound. For scale, human maintainers in the time-travel set modify tests in 9.9\% of compatibility fixes. The Claude Code systems do so in 38--53\% of apparent successes, and GPT-5.2 through Codex in 4\%.

Runtime enforcement shows that a low post-hoc source-only score can understate repair capability. Compared with full-patch scoring, GLM-5 drops by 21.8~pp (51.3\%$\to$29.5\%), while Kimi drops by only 3.1~pp (44.6\%$\to$41.5\%). Compared with the post-hoc source-only audit, however, both systems improve: Kimi rises from 22.8\% to 41.5\%, and GLM-5 from 24.4\% to 29.5\%. At repository level, 19 of GLM-5's 54 shortcut repositories and 24 of Kimi's 43 shortcut repositories pass once test writes are blocked. The same agent can make a source repair it skipped when test edits were available.
%Kimi's enforced 41.5\% is our clearest single-number capability anchor, 2.9$\times$ the pyupgrade~\cite{pyupgrade} baseline.

Manual inspection explains why post-hoc stripping is still needed. About 90\% of shortcut edits are plausible API adaptations inside tests, such as \texttt{nose}~$\to$~\texttt{pytest} rewrites; the remaining 10\% are direct bypasses such as \texttt{skip}/\texttt{xfail} injection or \texttt{assert} relaxation. In \texttt{cerberus}, for example, \texttt{pkg\_resources} appears in both source and tests. Under post-hoc scoring, agents migrate source to \texttt{importlib.metadata} and also edit the test; the audit then strips the test edit. Under enforcement, Kimi instead adds a small \texttt{pkg\_resources.py} compatibility shim that preserves the original test wording. The repair path depends on what the harness permits as much as on what the model can infer.

The source-only gap between GPT-5.2 through Codex and the Claude Code systems (49.7\% versus 19.7--24.4\%) remains a deployed-system observation. It may come from the underlying LLM, instruction following, patch granularity, framework behavior, prompt guardrails, or their interaction.
%A confirmatory GEE~\cite{liang1986gee} logistic regression (full table in artifact) preserves the cross-system pattern after controlling for $\log$ repo size and incompatibility type, with size dominating ($\log$\,Py\,files OR=0.47, $p<10^{-7}$). The regression supports the observed gap but does not isolate a cause.

\findingbox{Finding~RQ1: capability and compliance are different}{Full-patch pass rates combine source repair with test-edit shortcuts. Under source-only audit, 38--53\% of Claude Code apparent successes disappear, while GPT-5.2 through Codex uses such shortcuts in only 4\%; when test writes are blocked at runtime, Kimi still reaches 41.5\%. The evaluation regime changes what agents attempt, not only how we score them.}

% ------------------------------------------
% RQ2: Cross-system complementarity
% ------------------------------------------
\subsection{RQ2: Do agent systems solve the same repositories, or do their successes complement each other?}
\label{sec:complementarity}

The systems cover different parts of the repository set, and that spread remains after the source-only audit. Among the four Claude Code systems, the full-patch union is 110 of 193 repositories (57.0\%), above the best single-system rate of 51.3\% (GLM-5), while their intersection is only 55 of 193 (28.5\%). Adding GPT-5.2 through Codex raises the union to 121 of 193 (62.7\%) full-patch and 106 of 193 (54.9\%) source-only. Those are 10.9~pp full-patch and 5.2~pp source-only above that system alone, and 34.2~pp above the four-system intersection.

Two checks point in the same direction. Among pairs where both systems pass, file-level Jaccard on edited source files falls from 0.56 on Easy repositories to 0.43 on Medium repositories; the Hard tier has no both-passing pairs. Majority voting also loses to best-of-N, reaching only 45.1\% at a threshold of at least three of five systems. If successful repairs converged on the same edit locations, majority voting would lose less. In the 11 repositories where GPT-5.2 through Codex is the only successful system, trace inspection finds more coherent multi-file edits in its sessions, including parallel renames in at least four files. Claude Code sessions more often revert after intermediate test failures. Because model and framework are coupled, this is deployed-system trace evidence.

\findingbox{Finding~RQ2: systems cover complementary parts of the rescue set}{The five-system union reaches 54.9\% source-only, 5.2~pp above GPT-5.2 through Codex alone and 34.2~pp above the Claude Code intersection. Edited-file Jaccard among both-passing pairs falls from 0.56 to 0.43 as tasks get harder, so the complementarity reflects different repair profiles rather than sampling noise.}

% ------------------------------------------
% RQ3: Task structure + recurring failure patterns
% ------------------------------------------
\subsection{RQ3: What makes a rescue task hard for agents?}
\label{sec:task_structure}

Repository size and broad incompatibility labels explain only part of the difficulty. The sharper boundary appears when a patch has to keep assumptions consistent across files. We therefore classify each rescue hunk by the amount of coordination it requires. \textbf{L1} covers syntactic replacement, \textbf{L2} covers local single-file API adaptation, \textbf{L3} covers changes that propagate across files or dependency boundaries, and \textbf{L4} covers repairs where interacting components must be migrated together. The levels range from \texttt{typing.List}~$\to$~\texttt{list} at L1 and \texttt{inspect.getargspec}~$\to$~\texttt{getfullargspec} at L2, to NumPy~2.0 or \texttt{nose}~$\to$~\texttt{pytest} migrations at L3 and async or ABI refactoring at L4.
%Two annotators reach Cohen's $\kappa$~\cite{cohen1960kappa}~$=0.76$ over 4{,}334 paired hunks.
% For each repository, the \emph{success-anchored level} is the minimum reasoning ceiling among passing systems, or the maximum observed level if none passes.

Table~\ref{tab:reasoning_level} reports this analysis on the 116 hunk-labelled repositories that are also in the current 193-repository benchmark. Because this is a labelled subset, the table is not another full-benchmark total; it is a slice for asking whether success changes with the reasoning level of the patch. The main break is between local migration and whole-codebase coordination. L1 and L2 repairs are mostly routine (72--100\%). L3 keeps the cross-system spread seen in RQ2, with Sonnet at 66\% and GLM-5 at 92\%. At L4, where 14 labelled repositories require whole-codebase reasoning, GPT-5.2 through Codex is recorded as passing all 14 while every Claude Code system passes at most two.
Phase~2 preserves the historical test suite, so L4 often rewards repairs that keep the old API surface coherent under the new runtime, including compatibility shims and adapters. That is a reasonable target for abandoned dependents that still call the old interface, but it can also favor bridge-building over deeper redesign. The reasoning-level axis also tracks much of the incompatibility-type axis, with module removals usually falling into L1 or L2, NumPy and pandas churn into L3, and plugin or async refactoring into L4.

% Table 2: per-system pass rate by repository success-anchored reasoning level.
% Data source: annotation/final_labels.jsonl joined with eval_results, filtered to
% the current 193-repository valid benchmark from t1_verify_results.json.
% The unfiltered REPRO/out/level_x_system_success-anchored.csv includes 14
% labelled repos whose current T1 verification is PASS and must not be used for
% the paper table.
% Success-anchored level = min{ max_h-level(p) : p ∈ patches that PASSED on this repo },
% with fallback to overall max if no system passed. This represents the minimum
% reasoning ceiling sufficient for any deployed system to produce a Phase 2 PASS.
% A supplementary hunk-level distribution table (max-anchored) is provided in the appendix.
\begin{table}[t]
\centering
\caption{Python Phase~2 pass rate by success-anchored reasoning level on the 116 labelled repositories that are also in the current 193-repository benchmark. Cells show passing repositories with pass rate in parentheses; the row total appears in $n$. The table is a difficulty slice, not the full aggregate in Figure~\ref{fig:gap}. The short Codex column label denotes GPT-5.2 through Codex; Sonnet, GLM-5, Kimi, and MiniMax run in Claude Code.}
\label{tab:reasoning_level}
\small
\setlength{\tabcolsep}{1.8pt}
\renewcommand{\arraystretch}{1.05}
\begin{tabular*}{\columnwidth}{@{\extracolsep{\fill}}lrrrrrr@{}}
\toprule
\textbf{Level} & \textbf{$n$} & \textbf{Sonnet} & \textbf{GLM-5} & \textbf{Codex} & \textbf{Kimi} & \textbf{MiniMax} \\
\midrule
L1  &   4 & 4 (100) & 4 (100) & 4 (100) & 4 (100) & 4 (100) \\
L2  &  60 & 47 (78) & 54 (90) & 51 (85) & 47 (78) & 43 (72) \\
L3  &  38 & 25 (66) & 35 (92) & 28 (74) & 30 (79) & 23 (61) \\
L4  &  14 & 2 (14) & 2 (14) & \textbf{14 (100)} & 2 (14) & 0 (0) \\
\bottomrule
\end{tabular*}
\end{table}

The L4 pattern is clearest in \texttt{flexx}. Its patch has to align an event-loop migration (\texttt{asyncio.\allowbreak coroutine}~$\to$~\texttt{async~def}), a websocket layer ported away from the removed \texttt{asyncio.\allowbreak async}, and a JS-Python bridge that retypes message envelopes. GPT-5.2 through Codex produces a patch that compiles, imports, runs the round-trip suite, and survives the post-hoc audit. The Claude Code systems often identify correct local migrations, but the partial fixes do not compose. One repair breaks the bridge, while another preserves the bridge and misses the event loop. The failed traces therefore make coordination the main issue, even when API knowledge is present.

The pass-count tiers leave a smaller group of likely recoverable failures. We classify 65 repositories as Easy ($\geq$4 systems pass), 61 as Medium (1--3 pass), and 67 as Hard (none pass). Among the Hard repositories, 25 are near-misses with at least 95\% of tests passing; 7 fail on a single test, and 8 are hand-labelled as solvable by trivial L1 changes. These cases suggest 5--8\% headroom, although the trace sample cannot cleanly separate missing reasoning from premature termination. The rest of the Hard set is dominated by C-extension build failures and whole-codebase async refactoring.

The traces also show that some failures are about stopping as well as repair knowledge. Agents rarely ask for human input; most sessions, including failed ones, end with a self-declared completion. Three patterns recur, matching behavior reported for issue-driven repair~\cite{watanabe2025agenticcoding,jin2025systematicfailures,nashid2025beyondaccuracy}. In false-completion cases, final messages remain optimistic while visible test failures persist. A keyword detector flags 62--98\% of failed sessions per system, and a 30-session stratified validation sample gives 69\% precision and 95\% recall, for a precision-adjusted prevalence of 32--76\%. In regression cycles, 30\% of failed sessions either reach a clean intermediate test run or lose at least 20\% of the best in-session pass count by termination, which means agents often fail to return to the best state they have found. Effort--effectiveness inversion appears when turn count grows without improving outcomes. Session length ranges from 21 to 206 messages at the framework level with no positive correlation to success; within a system, failed sessions use 29--58\% more turns than successful ones ($p<0.01$). The framework-level length gap and the within-system failure gap are distinct mechanisms, even though both appear in turn count.

\findingbox{Finding~RQ3: hard rescues become a coordination cliff}{In the labelled Python slice, L1 and L2 repairs are routine, but L4 whole-codebase coordination separates systems: GPT-5.2 through Codex is recorded as passing all 14 L4 repositories, while Claude Code systems pass at most two. Part of this gap comes from the benchmark's preference for coherent adapters that preserve historical API surfaces. Failed traces add another failure mode: agents often stop confidently even when tests still fail, an intermediate clean state has been lost, or the failed run has already taken more turns than successful runs.}

% ------------------------------------------
% RQ4: Real-world value beyond Phase 2 PASS (PyCG + post-PASS validation)
% ------------------------------------------
% ==========================================
% RQ4: Case Study (now a research-question subsection inside Results)
% ==========================================
\subsection{RQ4: When the original tests pass, does the rescued library work in realistic use?}
\label{sec:case_study}

Phase~2 PASS establishes suite restoration, not practical reuse. For each Phase~2-passing unmaintained Python candidate, RQ4 asks three additional questions: whether the repaired package works in realistic entry-point scenarios, whether targeted bug-hunt probes reveal rescue-caused regressions, and whether the patch actually addresses the Python~3.13 compatibility failure rather than making an unrelated or unnecessary change.

We apply these checks to the 34 unmaintained Python candidates with a Phase~2 PASS. Five provide weak rescue evidence because the patch changes no source that could address the Python~3.13 break, adds an unrelated shim, or re-audit shows that the candidate already passed without rescue. Seven others pass Phase~2 but still fail an intended-use scenario. In these cases, a required dependency or service is absent, a clean install misses packaged data or a binary, the configured test scope misses the broken feature, or the main async and HTTP2 path still crashes. The remaining 22 work in realistic scenarios. Bug-hunt on those candidates finds five rescue-caused regressions, five upstream or pre-existing issues, and 12 rescues with no observed regression. Thus, of the 34 candidates that pass Phase~2, 22 pass realistic scenarios and 12 both pass bug-hunt and contain a patch that addresses the compatibility failure.

\begin{figure*}[t]
\centering
\includegraphics[width=0.8\linewidth]{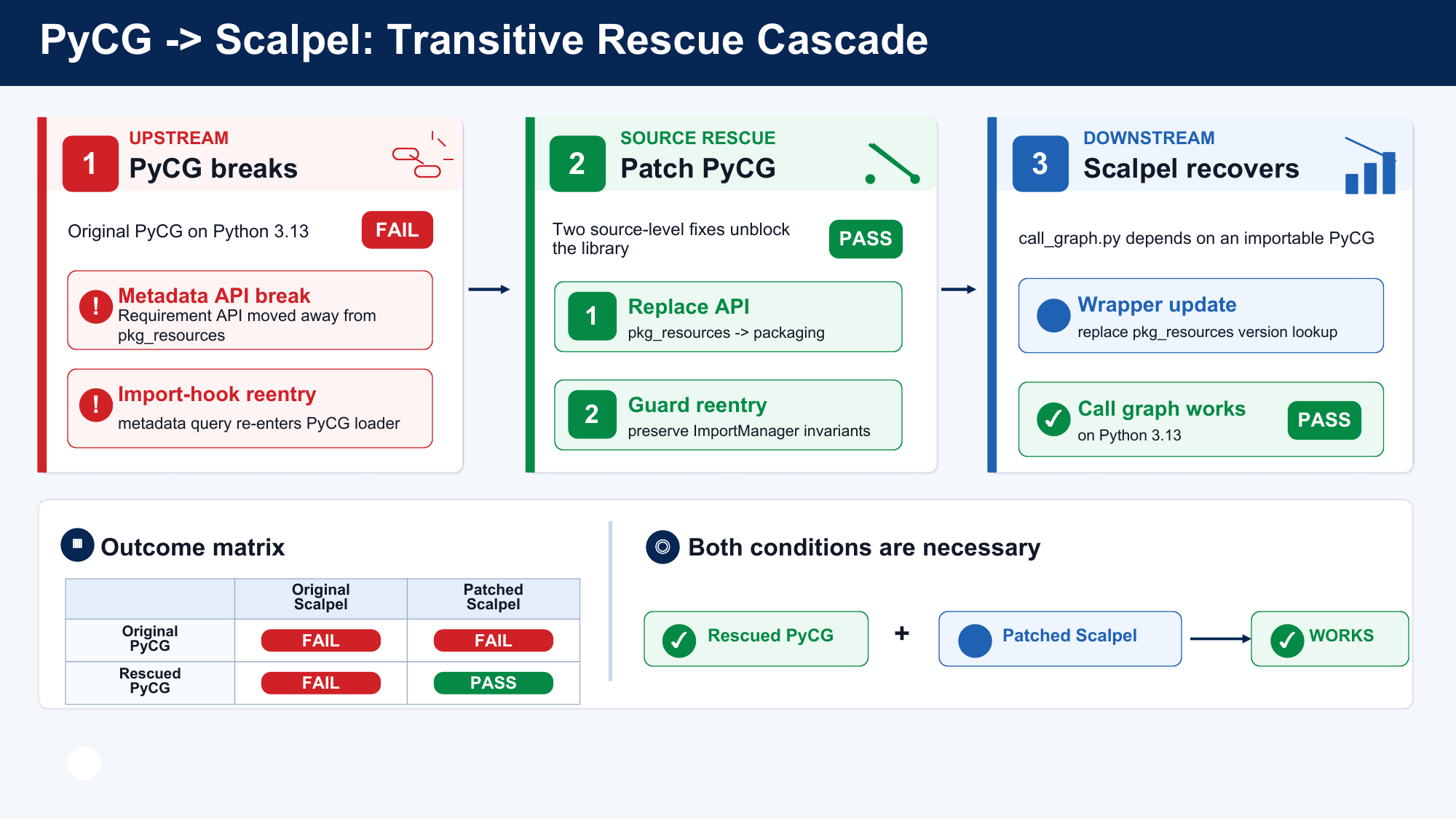}
\caption{PyCG $\to$ Scalpel: mechanism of a transitive rescue cascade. Region~1 shows the two-layer upstream failure in PyCG on Python~3.13 + \texttt{setuptools}~82. Region~2 shows the two source-level fixes that bring PyCG back. Region~3 shows the downstream Scalpel dependency and the small Scalpel-side compatibility edit.}
\label{fig:cascade}
\end{figure*}

We use PyCG~$\to$~Scalpel as a detailed case (Figure~\ref{fig:cascade}) because it shows why the additional checks matter: the historical suite passes after rescue, but the meaningful compatibility question appears only when the library is exercised through a modern downstream path. We then summarize the downstream patterns among rescues that pass realistic scenarios and the five rescue-caused regressions found by bug-hunt.

PyCG~\cite{pycg2021}, a Python call graph generator whose last commit was in November~2023, gives the clearest depth case. Its failure is a two-layer cascade, which puts it beyond a single syntactic replacement. The cascade generalizes beyond FastMCP~\cite{fastmcp}; it appears whenever a usage path exercises Python~3.13's lazy metadata loader while PyCG installs its path hook. FastMCP is a current example of the broader pattern, where an old library is wrapped behind a newer tool surface and the wrapper exposes a latent compatibility fault.

We wrap PyCG's call-graph generator behind a FastMCP~\cite{fastmcp} tool on Python~3.13 with \texttt{setuptools}~82, and the scenario exposes two failure layers (Figure~\ref{fig:cascade}, region~1). Layer~1 is a module-removal crash where the removed \texttt{pkg\_\allowbreak resources} module makes PyCG fail at import time before any of its code runs. Layer~2 appears only after Layer~1 is patched. \texttt{ImportManager.\allowbreak install\_hooks()} installs a custom path hook and then calls \texttt{invalidate\_\allowbreak caches()}. On Python~$\geq$~3.12, that call lazily loads \texttt{importlib.\allowbreak metadata} \emph{through} the just-installed hook before PyCG has set a current-module context. The bug lives at the intersection of new Python, old PyCG, and path hooks installed during normal usage, which helps explain why PyCG's 29 unit tests miss it.
Two repairs pass \texttt{validate.py}, a 3-file, 98-line reference fix written by an author and a 2-file, 95-line fix produced by GPT-5.2 through Codex. Both handle Layer~1 in the same way but choose different mechanisms for Layer~2 (Figure~\ref{fig:cascade}, region~2). The agent fix minimizes patch surface, while the reference fix also preempts a known Python~3.14 \texttt{ast} follow-up. Both produce the same FastMCP-path outcome, which shows that a rescue can be sufficient without covering every future deprecation.
%The reference fix was written after inspecting the GPT-5.2 through Codex trace, an anchoring risk we disclose even though the two Layer~2 mechanisms are architecturally different.

\label{sec:case_study:transitive}

A restored unit-test suite is strongest when the dependent code still works. The original unit tests may rarely exercise that boundary. PyCG makes this visible through \textsc{Scalpel}~\cite{scalpel2021}, a Python static analysis library that reuses PyCG for its \texttt{call\_graph} feature. Figure~\ref{fig:cascade}, region~3 shows that both sides must be adapted. The original Scalpel call-graph wrapper transitively crashes through PyCG; rescued PyCG plus a two-line Scalpel-side swap (\texttt{pkg\_\allowbreak resources.\allowbreak get\_\allowbreak distribution} $\to$ \texttt{importlib.\allowbreak metadata.\allowbreak version}) makes the harness pass. PyCG carries the harder share, a multi-file re-entrance bug versus a single-line downstream swap.

The four downstream cascades among the 22 realistic-scenario successes follow three patterns. In two cases, one upstream rescue plus a one-line downstream edit is enough (PyCG~$\to$~Scalpel; \texttt{flask-restful}~$\to$~swagger). In another, the upstream rescue provides compatibility shims that the downstream reaches only through restricted API paths (\texttt{pyasn1-modules}~$\to$~\texttt{cryptography}, partial cascade). In the fourth, upstream and downstream patches touch different Python~3.13 break surfaces but still compose cleanly (\texttt{pymorphy2}~$\to$~\texttt{yargy}). The Layer~2 re-entrance bug also shows why rule-based modernizers miss some cross-version runtime interactions, because the failure depends on the execution order between Python's lazy metadata loader and PyCG's path-hook installation.

\label{sec:case_study:generalised}

The same checks apply beyond PyCG. Four rescues that pass realistic scenarios directly unblock a live downstream consumer on Python~3.13 (PyCG~$\to$~Scalpel, \texttt{pymorphy2}~$\to$~\texttt{yargy}, \texttt{pyasn1-\allowbreak modules}~$\to$~\texttt{cryptography}, \texttt{flask-\allowbreak restful}~$\to$~\texttt{swagger-3}), and four can be exposed as FastMCP-style agent tools. Full per-repository traces are in the artifact. As in the aggregate counts, a restored test suite is useful evidence, while scenario use and uncovered regressions still require separate checks.

\textbf{Bug-hunt summary.} We applied bug-hunt to every rescue that passed the realistic scenarios. Five rescues exhibit rescue-caused regressions missed by the original suite, including blanket exception handlers, silently mangled merge markers, lost subprocess history, stdlib-method shadowing, and dropped failing-URL exception paths. Full per-rescue diffs are in the artifact. Regression-shaped issues that trace to upstream or pre-existing causes are excluded from the rescue-caused count.

\findingbox{Finding~RQ4: practical reuse needs checks beyond the historical suite}{Among 34 unmaintained Python candidates that pass Phase~2, 22 work in realistic scenarios and 12 pass bug-hunt with patches that address the compatibility failure. Seven still fail intended-use scenarios, five introduce uncovered regressions, and five pass Phase~2 without a meaningful compatibility repair. Four downstream cascades show why this matters: rescued libraries are often reused through dependents and tool wrappers that the original suite never exercises.}

% ------------------------------------------
% Java cross-ecosystem study
% ------------------------------------------
\subsection{Cross-ecosystem extension: what Java separates}
\label{sec:java_results}

The four RQs are evaluated primarily on Python. We use Java as a cross-ecosystem extension of RQ1 and RQ2, not as a fifth research question or a direct rate comparison. We run three systems on the 122 unmaintained Java repositories of \S\ref{sec:dataset}: GPT-5.2 through Codex, GLM-5 through Claude Code, and Kimi~K2.5 through Claude Code. Because static typing separates compilation failures from runtime failures, Java lets us ask whether shortcut behavior and complementarity persist when the failure type is more observable.

\begin{table}[t]
\centering
\caption{Java rescue results (JDK~21) by Phase~1 failure type. The short Codex column label denotes GPT-5.2 through Codex; GLM-5 and Kimi run in Claude Code. Cells show pass count and pass rate except retain rows, where Retain means source-only relative to full-patch.}
\label{tab:java_results}
\small
\setlength{\tabcolsep}{2.5pt}
\renewcommand{\arraystretch}{1.04}
\begin{tabular*}{\columnwidth}{@{\extracolsep{\fill}}lrrrrr@{}}
\toprule
\textbf{Failure} & \textbf{$n$} & \textbf{Codex} & \textbf{GLM-5} & \textbf{Kimi} & \textbf{Union} \\
\midrule
Compile & 52 & 45 (86.5) & 33 (63.5) & 45 (86.5) & 49 (94.2) \\
Runtime or test & 70 & 46 (65.7) & 33 (47.1) & 57 (81.4) & 59 (84.3) \\
\midrule
Overall, full & 122 & 91 (74.6) & 66 (54.1) & 102 (83.6) & 108 (88.5) \\
Overall, source & 122 & 87 (71.3) & 58 (47.5) & 76 (62.3) & 95 (77.9) \\
Retain, overall & -- & 96\% & 88\% & 75\% & 88\% \\
\midrule
Retain, compile & -- & 84\% & 76\% & 73\% & -- \\
Retain, runtime & -- & \textbf{107\%} & 100\% & 75\% & -- \\
\bottomrule
\end{tabular*}
\end{table}

\textbf{When test edits hurt.} Java makes one Python-invisible effect visible: test edits can damage an otherwise working source repair. On six Java repositories, test modifications from GPT-5.2 through Codex introduced compile or runtime errors, and stripping those edits \emph{restored} Phase~2 PASS. This pushes the system's Runtime retain to 107\% (Table~\ref{tab:java_results}). GPT-5.2 through Codex and GLM-5 through Claude Code retain near-perfectly on runtime failures (107\%, 100\%) but drop on compile failures (84\%, 76\%); 21 of 122 (17\%) Java outcomes from GPT-5.2 through Codex are perturbed by test edits in either direction. The same data suggests a system-level policy difference: GPT-5.2 through Codex and GLM-5 through Claude Code modify tests selectively, mostly when no source compile fix is available, while Kimi~K2.5 through Claude Code uses them as a general tactic. All 7 of Kimi~K2.5 through Claude Code's \emph{structural} test edits collapse under source-only audit.

Complementarity carries over to Java. The three-system Java union reaches 88.5\% full-patch and 77.9\% source-only, 6.6~pp above the best single system (GPT-5.2 through Codex, 71.3\% source-only). This gap is slightly larger than the within-Python five-system source-only gap of 5.2~pp. GPT-5.2 through Codex contributes the most unique source-only solves (13) despite Kimi~K2.5 through Claude Code's higher full-patch rate, paralleling the divergence between Codex and Claude Code in \S\ref{sec:complementarity}.

Java pass rates are higher than Python (three-system union 88.5\% versus five-system union 62.7\%), but several protocol differences shape that comparison. The Java protocol includes pre-admission \texttt{pom.xml} normalization, uses an unmaintained-only dataset, lacks an enforced ablation, and benefits from sharper compile-time error signals and more standardized tooling. We therefore report the gap qualitatively and leave Java post-PASS validation as future work.

\findingbox{Finding~Java: static typing exposes shortcut harm}{Java reveals an effect hidden in Python: in 6 repositories, stripping test edits from GPT-5.2 through Codex restores Phase~2 PASS, producing 107\% runtime retain. Complementarity also transfers: the three-system union reaches 77.9\% source-only, 6.6~pp above that system alone. The higher absolute Java pass rate remains qualitative because dataset and protocol confounds are substantial.}

% Sec 5: Related Work
% ==========================================
% Related Work
% ==========================================
\section{Related Work}
\label{sec:related}

\reporescue{} is closest to LLM-agent repair and code-migration benchmarks, but differs in both the task input and the evidence required for success. Repair benchmarks usually start from an issue, a failing test, or some fault-localization signal; migration benchmarks often constrain the change to a known API, library, or target version. \reporescue{} starts from a whole repository whose suite fails after modernization, and the agent receives the failing state without a supplied symptom or location. It also treats a green test suite as the first signal: source-only auditing, runtime enforcement, and post-PASS validation separate source repair from test-edit shortcuts and from usability beyond the historical suite. We organize adjacent work into two strands: repair and migration benchmarks, and behavioral or ecosystem studies.

\subsection{Repair and Migration Benchmarks}
SWE-bench~\cite{jimenez2024swebench} and agents built on it (SWE-agent~\cite{yang2024sweagent}, Agentless~\cite{xia2024agentless}) supply an issue and evaluate against its target test. Localized repair systems (RepairAgent~\cite{bouzenia2025repairagent}, UniDebugger~\cite{lee2025unidebugger}, TSAPR~\cite{hu2025aprmcts}, CodeAgent~\cite{zhang2024codeagent}, HAFixAgent~\cite{chen2025hafixagent}, DynaFix~\cite{huang2025dynafix}, RGFL~\cite{sepidband2026rgfl}) likewise assume a failing test and often fault localization; recent repository-scale work (Chen et al.~\cite{mu2025repolevelapr}, RepoRepair~\cite{li2026reporepair}, SgAgent~\cite{zhang2026sgagent}, Yang et al.~\cite{yang2025repokg}, RepoAI~\cite{chondamrongkul2026repoai}, RepoAudit~\cite{guo2025repoaudit}) broadens the spatial scope while keeping a supplied symptom or location. Chen et al.~\cite{mu2025repolevelapr} is closest in scope, but still starts from a localized bug report tied to a specific failing test; \reporescue{} starts from suite-wide failure at collection time and scores source repair despite root causes spanning dependency manifests and stdlib removals.

Function-level benchmarks (Almeida et al.~\cite{almeida2025copilot}, CodeMEnv~\cite{cheng2025codemenv}, GitChameleon~\cite{misra2025gitchameleon2}) and library-migration tools (PCART~\cite{wang2024pcart}, MigrateLib~\cite{islam2025migratelib}, PyMigBench~\cite{islam2023pymigbench}) bound the change to a single API or library. FreshBrew~\cite{joshi2025freshbrew} is closest in spirit: 228 Java projects to JDK~17 with a coverage-preservation guard. \reporescue{} adds an enforced-runtime ablation and a per-repository test-count guard, so it can distinguish final-patch shortcuts, shortcut attempts during the run, and test deletion.

\subsection{Behavioral and Ecosystem Studies}
ExecutionAgent~\cite{bouzenia2025executionagent} studies how agents \emph{build} arbitrary projects. Software-aging and abandonment studies explain why useful projects lose maintainers~\cite{coelho2017why,avelino2019abandonment,valiev2018ecosystem}. Dependency-supply-chain studies measure the other side of the same pressure: libraries lag behind current releases~\cite{cox2015dependencyfreshness,decan2018technicallag}, developers often delay dependency updates~\cite{kula2018developers}, package ecosystems propagate churn and vulnerabilities through dependency networks~\cite{decan2019empirical,liu2022demystifying}, and build reproducibility can fail when environments drift~\cite{mukherjee2021fixing,bartlett2025pllm,vangala2025reproducibility}. \reporescue{} turns these ecosystem observations into executable rescue tasks: each subject is admitted only after the old environment passes, the modern environment fails, and the same historical suite can judge a source-only adaptation.

Methodologically closest are recent agent-behavior studies (Watanabe et al.~\cite{watanabe2025agenticcoding}, Jin and Chen~\cite{jin2025systematicfailures}, Nashid et al.~\cite{nashid2025beyondaccuracy}, Zhu et al.~\cite{zhu2026agentframeworkbugs}) that look beyond final PASS or FAIL outcomes at trajectory dynamics. We extend that line to a whole-repository no-issue setting and use the contrast between post-hoc auditing and runtime enforcement as a second way to observe shortcut behavior.

% Sec 6: Discussion
% ==========================================
% Discussion
% ==========================================
\section{Discussion}
\label{sec:discussion}

\textbf{What the benchmark makes measurable.} \reporescue{} turns a common maintenance problem into an observable agent task. The task starts from code that once worked, gives the agent a failing modern environment with no issue report or fault localization, and asks the agent to recover compatibility under a source-only constraint. This matters because many compatibility breaks combine runtime drift, dependency churn, stale tests, and downstream reuse; they rarely reduce to a single API replacement. A dataset with Phase~0, Phase~1, and Phase~2 validation can separate those parts instead of treating every green suite as the same kind of success.

\textbf{What the findings say about agents.} The results support treating deployed agents as model--framework systems. Closing the test-edit shortcut changes Kimi's observed behavior, so compliance is part of capability rather than a post-processing detail. The union-vs.-single gap shows that systems fail on different repositories, and the L4 cliff shows where that difference becomes most visible: coordinated edits that preserve an old API surface across interacting files. The reasoning-level hierarchy therefore gives a way to state capability claims at the granularity where the failures actually separate.

\textbf{How the results can be used.} The larger use case is to make dependency paths movable again, because a single abandoned library can block a modernized application or tool. Modernization failures often surface at the top-level project, while the hard break may sit in a dormant transitive dependency. Common cases include an import that no longer loads, missing packaged data, an old API still called by downstream code, or an intermediate library that cannot be wrapped as an agent tool. \reporescue{} treats each such library as a repairable link. When the broken links along a supply chain are rescued under source-only constraints and then checked through downstream scenarios, the path has a route to a modern runtime without replacing every package at once.
This view suggests a practical workflow. Benchmark designers should keep post-hoc source-only scoring, runtime enforcement, and post-PASS validation separate, because they answer different questions. Tool builders can route L4-like tasks to systems that handle cross-file coordination better and add in-flight checks for false completion or regression cycles. Maintainers can rank abandoned libraries by downstream reach, rescue the load-bearing intermediate packages first, and validate them through real dependents or tool wrappers. The PyCG~$\to$~Scalpel cascade illustrates the pattern: once PyCG's deeper failure is repaired, the downstream Scalpel-side edit is small, and the combined path works on Python~3.13. This remains path-level evidence, with downstream execution and regression probes strengthening what the historical suite can show.

% Sec 7: Threats
% ==========================================
% Threats to Validity
% ==========================================
\section{Threats to Validity}
\label{sec:threats}

Following standard validity categories for empirical software engineering~\cite{runeson2009guidelines,kitchenham2002preliminary}, the following threats bound how far the behavioral claims should be read: \reporescue{} is the instrument through which we observe agent behavior, not a calibrated estimator of underlying capability.

\textbf{Construct and comparison.} Rates describe what systems did under our protocol, not what they could do under different prompting or sampling. Single-trial-per-cell design gives union and intersection counts $\pm$5--10 repository drift; qualitative regularities are robust to it. The four Claude Code systems share infrastructure, while GPT-5.2 through Codex differs in both LLM and framework, so we report it as a separate observation; provider-side sampling defaults remain a residual confound on the within-framework spread. The L1--L4 hierarchy is also a construct: two annotators reach Cohen's $\kappa$~\cite{cohen1960kappa}~$=0.76$ (per-system 0.70--0.81), with $\sim$half the disagreement at the L2 versus L3 boundary. Table~\ref{tab:reasoning_level} therefore uses only the 116 repositories that are both hunk-labelled and present in the current benchmark, and we use the hierarchy for the L4-cliff finding rather than fine L2 versus L3 distinctions.

\textbf{Dataset and harness integrity.} Python combines 47 unmaintained repositories and 146 time-travel snapshots; \texttt{repo\_type} is not significant in our GEE~\cite{liang1986gee} regression after controlling for size, system, and incompatibility type ($p$=0.75), and unmaintained-only rates equal or exceed combined rates, the opposite of a memorisation explanation. Claims specific to unmaintained projects (RQ4) use the 47-repository subset only. The Python Phase~2 protocol preserves the historical test command verbatim; auditing 965 RQ1 trial logs found 6 of 436 (1.4\%) PASS outcomes with a ``no tests ran'' signal concentrated in \texttt{wssh}, bounding aggregate pass-rate inflation by 1.4~pp, with per-trial flags shipped in the artifact.

\textbf{External validity.} Python's dynamic typing creates a specific breakage profile; qualitative patterns recur on Java despite the asymmetric design, but rate comparisons partly reflect dataset composition. Forbidding dependency changes abstracts away a channel real maintainers use (29.8\% of time-travel fixes); this isolates source reasoning rather than claiming source-only is the realistic deployment mode. RQ4 adds scenario and bug-hunt probes because Phase~2 PASS establishes suite restoration, not semantic correctness.

% ==========================================
% 8. Conclusion
% ==========================================
\section{Conclusion}
\label{sec:conclusion}

\reporescue{} starts from a common maintenance problem: useful repositories can outlive the environments that made them work. We turn that problem into a benchmark by requiring each subject to pass in its historical environment, fail after modernization, and be rescued through source-code changes under the original test command. This construction lets the study ask whether agents can make old software usable again and what kind of repair a restored test suite actually represents.
Agents can restore many compatibility failures, and different systems cover complementary repositories. At the same time, full-patch success often hides test-edit shortcuts, whole-codebase coordination remains a sharp difficulty boundary, and passing the historical suite does not by itself establish practical reuse. The broader lesson is that compatibility rescue needs an evaluation stack: source-only scoring and runtime enforcement to separate source repair from shortcuts, reasoning-level analysis to locate coordination limits, and scenario or regression checks to decide whether a rescued library can be used again.

\bibliographystyle{IEEEtran}
\bibliography{main}

\end{document}